\documentclass[12pt]{article}
\usepackage[english]{babel}
\usepackage{amsmath,amssymb,amsbsy,amstext,amsthm,booktabs,simplewick,exscale,relsize,slashed,graphicx,amsfonts,upgreek,cite}
\usepackage{multirow,dcolumn,bm,enumerate,url,hyperref,placeins}
\usepackage{color}
\usepackage[colorinlistoftodos]{todonotes}

\newcommand{\mchib}{\overline{m}_\chi}

\title{\bf Detecting Composite Dark Matter with Long Range and Contact Interactions in Gas Clouds}

\author{Amit Bhoonah$^\eth$, Joseph Bramante$^{\eth,\dagger}$, Sarah Schon$^{\eth,\dagger}$, Ningqiang Song$^{\eth,\dagger}$\\
{\small $^\eth$ The Arthur B. McDonald Canadian Astroparticle Physics Research Institute and} \\ {\small Department of Physics, Engineering Physics, and Astronomy,} \\ {\small Queen's University, Kingston, Ontario, K7L 2S8, Canada}\\ {\small $^\dagger$Perimeter Institute for Theoretical Physics, Waterloo, Ontario, N2L 2Y5, Canada}}

\begin{document}
\maketitle

\begin{abstract}
Cold interstellar gas clouds provide an exciting new method to discover dark matter. Their immense size makes them uniquely sensitive to interactions from the heaviest, most rarefied dark matter models. Using gas cloud observations, we derive constraints on heavy composite dark matter coupled to the Standard Model through a light dark photon for dark matter up to a thousand solar masses. We find gas clouds are also sensitive to very large composite dark matter that interacts with nuclei through a fixed contact interaction cross-section. We also study the contact interaction model and implement multiscatter and overburden analyses to obtain bounds from experiments like CDMS, CRESST, DAMA, XQC, and XENON1T.
\end{abstract}

\section{Introduction}
Cosmological observations and galactic dynamics have established that dark matter provides the bulk of matter in our universe. As a consequence, it is imperative that we determine the nature of dark matter as we advance knowledge of fundamental physics and the beginnings of the universe. Rather little is known about dark matter's mass and its couplings to known particles, and so we must develop new methods to unmask all plausible dark matter candidates. 

Although there is a vast body of literature investigating weakly interacting particle dark matter (aka WIMPs) with a mass near that of the proton, comparatively less is known about heavier dark matter. Nevertheless, it has been clear since the first proposal of quark droplet dark matter \cite{Witten:1984rs}, and more recent composite dark matter models \cite{Bagnasco:1993st,Zhitnitsky:2002qa,Foadi:2008qv,Kribs:2009fy,Wise:2014ola,Wise:2014jva,Hardy:2014mqa,Detmold:2014qqa,Detmold:2014kba,Krnjaic:2014xza,Jacobs:2014yca,Gresham:2017zqi,Gresham:2017cvl,Ge:2017idw,Bai:2018dxf,Bai:2019ogh}, that dark matter may be a rather heavy composite state. Despite the existence of these heavy dark matter models, it has been an ongoing challenge to obtain sensitivity to dark matter heavier than the Planck mass, because the Planck mass is the maximum mass a human-scale dark matter detector can find in a background-free runtime of about a year \cite{Bramante:2018qbc,Bramante:2018tos}. In this paper we will find that cold gas clouds in our Milky Way galaxy provide unparalleled sensitivity to composite dark matter models with masses ranging from the mass of a proton, up to many times the mass of the sun. Prior publications have studied interstellar gas clouds as dark matter detectors for strongly interacting, millicharged, and ultralight dark photon dark matter models \cite{Bhoonah:2018wmw,Bhoonah:2018gjb,Chivukula:1989cc}. There have also been recent proposals to use the cold and hot phases of interstellar gas in dwarf galaxies to look for dark matter, and in particular primordial black hole dark matter \cite{Wadekar:2019xnf,Lu:2020bmd,Kim:2020ngi}. Here we will focus on how cold gas clouds near the Milky Way galactic center can provide leading sensitivity to models of composite dark matter.

The remainder of this paper proceeds as follows. In Section \ref{sec:ac} we find gas cloud sensitivity for composite asymmetric dark matter which interacts with Standard Model particles through a long-range dark photon mediator. In Section \ref{sec:contact}, we detail gas cloud sensitivity for an effective model of composite dark matter with a fixed contact interaction cross-section for scattering with nuclei; bounds on this model are also obtained from terrestrial experiments, including CDMS, CRESST, DAMA, XQC and XENON1T. In Section \ref{sec:conc} we conclude. In Appendix \ref{sec:appbounds}, we present a summary of bounds on composite dark matter's contact interactions from a search for tracks in mica and Skylab's plastic etch detectors.

\section{Composite Asymmetric Dark Matter}
\label{sec:ac}
Interstellar gas clouds have been observed with temperatures around 100 Kelvin and lengths spanning tens of parsecs \cite{McClure-Griffiths:2013awa}. These characteristics result in exquisite gas cloud sensitivity to very heavy composite dark matter that interacts primarily through long-range forces. Here we will focus on explicit models of composite bound states detailed in \cite{Wise:2014ola,Wise:2014jva,Hardy:2014mqa,Gresham:2017zqi,Gresham:2017cvl}, where the composites are comprised of asymmetric dark matter constituents bound together by a scalar field. For a review of asymmetric dark matter, see $e.g.$~\cite{Zurek:2013wia,Petraki:2013wwa,Kaplan:2009ag,Lin:2011gj}. A sufficiently strong and long-range force between asymmetric fermion dark matter will result in many dark fermions being bound together after an early period of ``dark nucleosynthesis.'' As more constituents are added, the combined state of dark fermions will saturate to a constant density once the Fermi degeneracy pressure dominates the state. The terrestrial direct detection prospects for these asymmetric composite states (dubbed ADM nuggets) have been recently studied in Ref.~\cite{Coskuner:2018are}. Although on-going and future underground low threshold direct detection experiments can potentially put quite stringent bounds on nugget masses below $10^{17}$~GeV, their sensitivities are lost quickly for larger masses due to intrinsic flux sensitivity problems along with the overburden problem, where the expected interaction can be so strong enough that nuggets are slowed to small speeds by scattering against the Earth's crust before they reach the detector. These nuggets fail to trigger the detector with their small thermal velocities. One advantage of gas clouds as dark matter calorimetric detectors is that they are sensitive to nuggets with masses below $\sim 10^{60}$~GeV~\cite{Bhoonah:2018gjb} as a consequence of the extremely low baryon number density.

The ADM nugget can be parameterized by two quantities, the reduced constituent mass $\mchib$ which incorporates the in-medium contribution to the bare constituent mass, and the total nugget mass $M_X$~\cite{Coskuner:2018are}. A nugget with $N_X=M_X/\mchib$ constituents has a radius given by 
\begin{equation}
    R_X=\left(\dfrac{9\pi}{4}\dfrac{M_X}{\mchib^4}\right)^{1/3}\,,
    \label{eq:RX}
\end{equation}
assuming a saturated constituent density inside the nugget. With the nugget so defined, we turn do dark matter interactions with Standard Model particles. We will be primarily interested in the interaction cross-section between an ADM nugget and a Standard Model (SM) target due to a long range interaction,
\begin{equation}
    \dfrac{d\sigma}{dq}=\dfrac{q}{2v^2\mu_{Xt}^2}\widetilde{\sigma}(q)N_X^2|F_X(q)|^2\,,
    \label{eq:dsigmadq}
\end{equation}
where $\mu_{Xt}$ is the reduced mass between the nugget and the target electron or nucleon, $q$ denotes the momentum transfer and $\widetilde{\sigma}$ reads
% \begin{equation}
%     \widetilde{\sigma}=\dfrac{g_t^2g_\chi^2}{4\pi}\dfrac{\mu_{Xt}^2}{(q^2+m_{A'}^2)^2}\,.
% \end{equation}
\begin{equation}
    \widetilde{\sigma}=\dfrac{4\pi\epsilon^2\alpha_\mathrm{EM}\alpha_X\mu_{Xt}^2}{(q^2+m_{A'}^2)^2}\,.
\end{equation}
We assume the nugget-SM interaction is dominated by the exchange of a dark photon mediator $A'$ which kinetically mixes with SM photon with the mixing parameter $\epsilon$. Here $\alpha_\mathrm{EM}\simeq 1/137$ is the fine structure constant and $\alpha_X=g_\chi^2/(4\pi)$ defines the gauge coupling between the dark photon and nugget constituent $\chi$. 

Because the vector $A'$ will couple to $X$ particles, and provide an additional repulsive force between them, a few comments are in order about the validity of \eqref{eq:RX}, and the stability of very large dark matter composites under the influence of a vector mediator. In the Standard Model, the presence of a vector force can destabilize nuclei and specifically prevents the formation of large nuclei. This is related to the fact that the binding potential of quantum chromodynamics is limited by the QCD confinement scale to lengths $\lesssim {\rm fm}$. In the case of ADM nuggets bound together by a scalar field, these nuggest can remain stable up to very large masses \cite{Gresham:2017zqi}, so long as this scalar field is less massive and couples more strongly than the vector mediator $A'$, i.e. $g_\phi^2/m_\phi \gg g_\chi^2/m_{A'}^2$ with $g_\phi$ ($g_\chi$) and $m_\phi$ ($m_{A'}$) the coupling and mass of the scalar (vector) mediator. We will always assume this regime, where the nugget binding energy is large enough to overcome the repulsive force introduced by the vector mediator and the nugget can become arbitrarily large -- we consider nugget masses ranging up to hundreds of solar masses. As the constituents of nuggets can be modeled as relativistic fermi gas, fermi pressure will prevent the nuggets from collapse. Consequently fine tuning of the vector force is not required to balance the attractive potential of the scalar mediator. However, the vector mediator tends to affect the saturation density of nuggets. We refer readers to~\cite{Gresham:2017zqi} for detailed discussions.

If scattering with SM nuclei, an additional factor $Z^2|F_N(q)|^2$ needs to be introduced to Eq.~\eqref{eq:dsigmadq} where the Helm nucleus form factor is given by
\begin{equation}
    F_N(q)=\dfrac{3j_1(qR_N)}{qR_N}e^{-q^2s^2/2}\,, 
\end{equation}
where $s=0.9$~fm and the nucleus radius is roughly $R_N\simeq1.2\,\,\mathrm{fm}\times A^{1/3}$, and $j_1$ is the Bessel function of the first kind. Since the ADM nugget has a finite radius $R_X$, a nugget form factor is also required to account for the mass distribution of the nugget, which we take to be
\begin{equation}
    F_X(q)=\dfrac{3j_1(qR_X)}{qR_X}\,.
\end{equation}

When traversing the gas cloud, the ADM nugget will heat the gas cloud by scattering with the nuclei and free electrons in the gas cloud. The energy deposited per scatter is $q^2/2m_N$ for nuclear scattering or $q^2/2m_e$ for electron scattering. In the case that nuclear scattering dominates DM-gas cloud energy transfer, the dark matter gas cloud volumetric heating rate (VDHR) is given by
\begin{equation}
    VDHR=\sum\limits_i\dfrac{\rho_X}{M_X}f_i\dfrac{m_n n_b}{m_{N_i}}\dfrac{1}{2\mu_{Xn}^2}\int\dfrac{dv}{v}f(v)\int_{q_{\min}}^{q_{\max}} dqq\dfrac{q^2}{2m_{N_i}}\dfrac{d\sigma_i}{dq}\,,
    \label{eq:heatingratenucleon}
\end{equation}
In Eq.~\eqref{eq:heatingratenucleon} we also sum up the contribution from dark matter scattering with different nuclei. The predominant elements in gas clouds are hydrogen, helium, oxygen and iron with mass fractions 
\begin{equation}
    \{f_H,f_{He},f_{O},f_{C},f_{Fe}\}=\{0.71,0.27,0.01,0.004,0.0014\},
\end{equation} 
where these relative abundances assume a solar metallicity. Dark matter's volumetric heating of the gas cloud can also be predominantly from the ADM nugget's interaction with electrons, in which case
\begin{equation}
    VDHR_e=\dfrac{\rho_X}{M_X}n_e\dfrac{1}{2\mu_{Xe}^2}\int\dfrac{dv}{v}f(v)\int_{q_{\min}}^{q_{\max}} dqq\dfrac{q^2}{2m_e}\dfrac{d\sigma_e}{dq}\,.
    \label{eq:heatingrateelectron}
\end{equation}

Here we will set bounds using gas cloud G357.8-4.7-55~\cite{McClure-Griffiths:2013awa}, which is characterized by a temperature $T_g=137$~K and a baryon number density $n_b = 0.42$~cm$^3$, and a free electron number density $n_e=10^{-3}$~cm$^{-3}$. The dark matter density near the gas cloud is approximately 17~GeV/cm$^3$, assuming standard DM MW halo density distributions \cite{Bhoonah:2018gjb}. 
We take the dark matter velocity in the rest frame of the gas cloud to follow a Maxwellian distribution where we assume a characteristic velocity $v_0=220$~km/s and a relative gas cloud velocity $v_{gc}=180$~km/s, as detailed in \cite{Bhoonah:2018gjb}. In order for the gas cloud to cool to the temperature observed today, we require the dark matter heating rate not exceed the volumetric cooling rate (VCR) today, ie
\begin{equation}
    VDHR\leq VCR
\end{equation}
where the average cooling rate is estimated to be $3.4\times10^{-28}$~erg~cm$^{-3}$s$^{-1}$~\cite{Bhoonah:2018gjb}.

\begin{figure}
    \centering
    \includegraphics[width=0.49\textwidth]{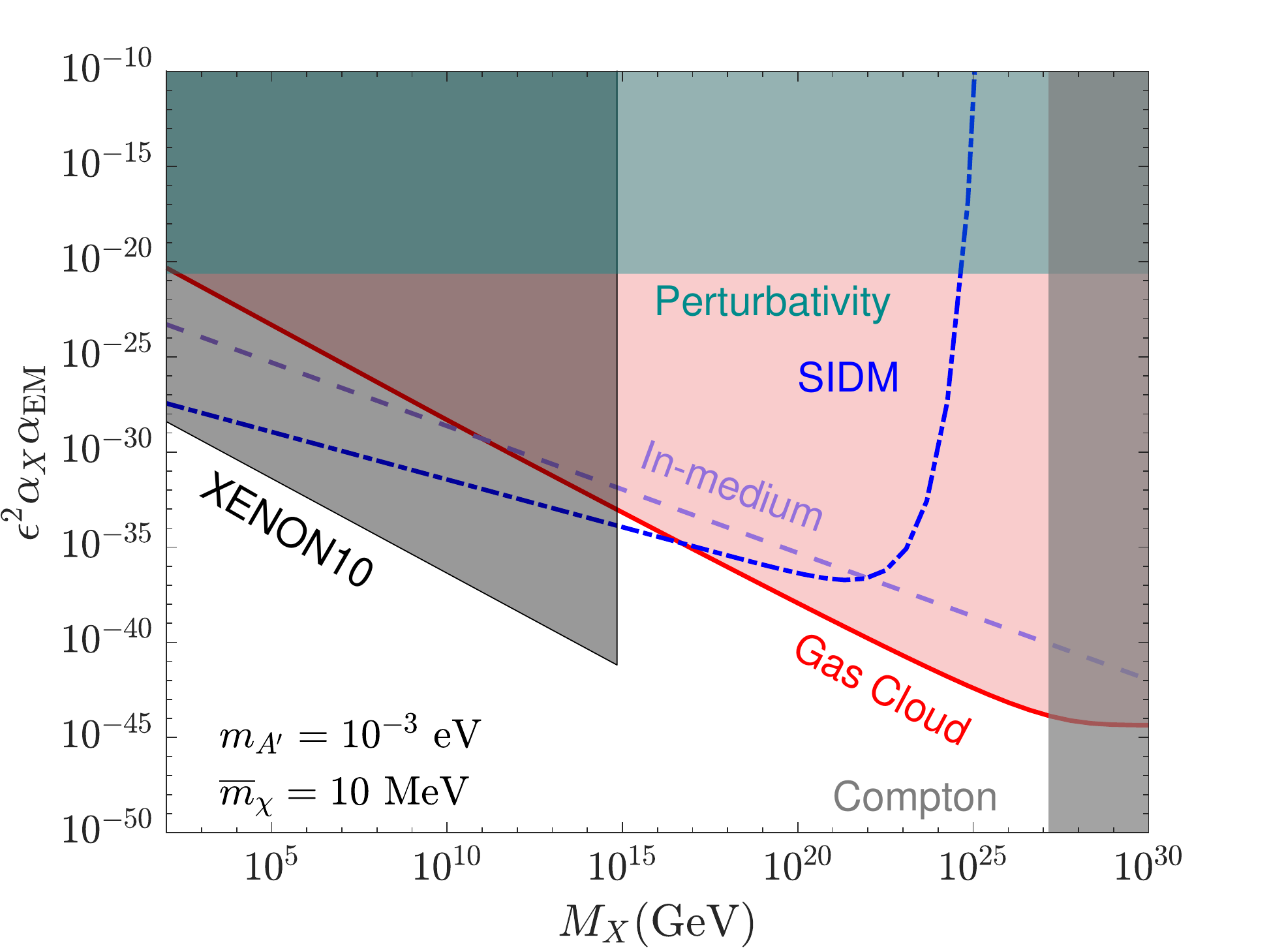}
    \includegraphics[width=0.49\textwidth]{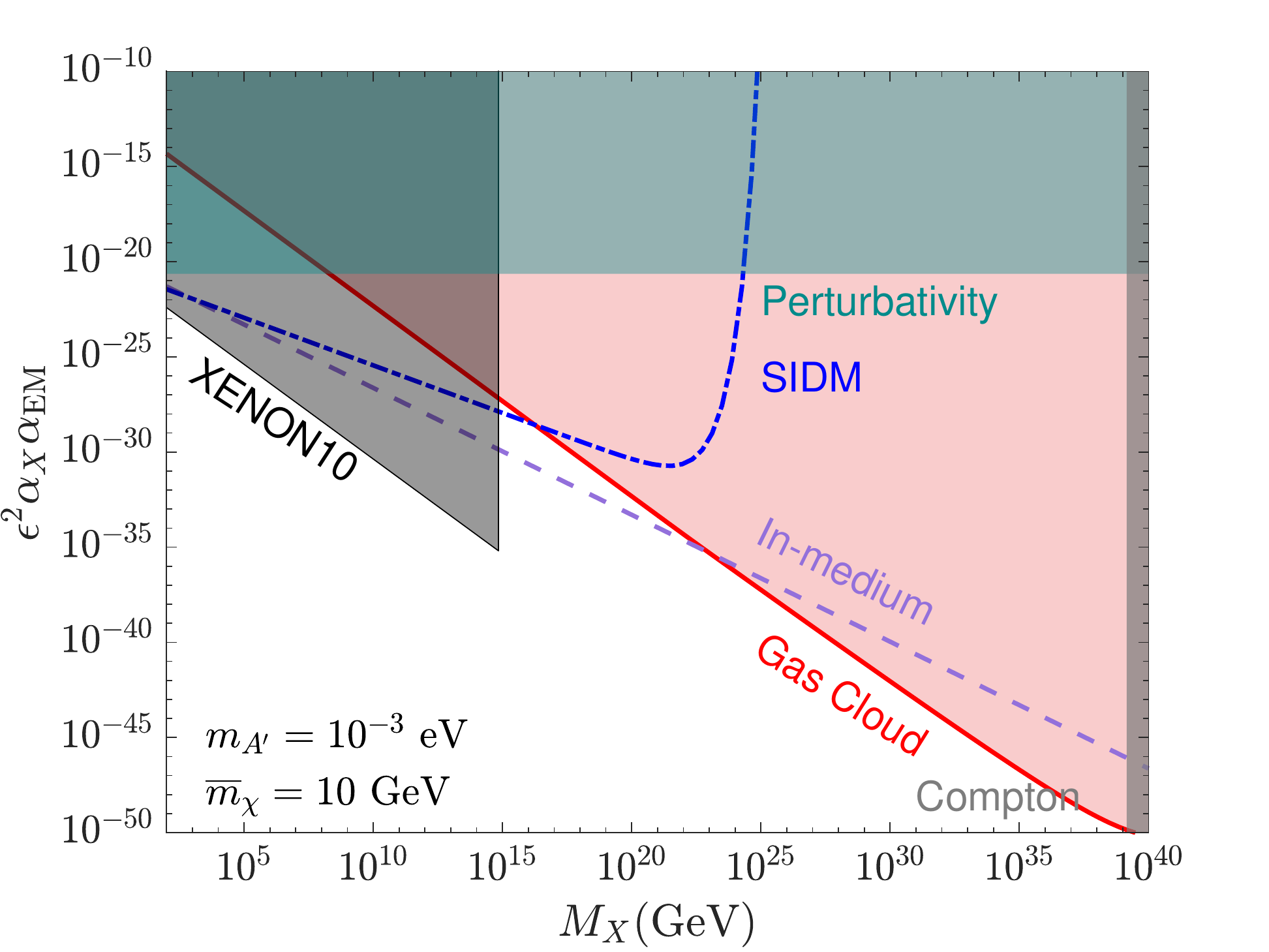}\\
    \includegraphics[width=0.49\textwidth]{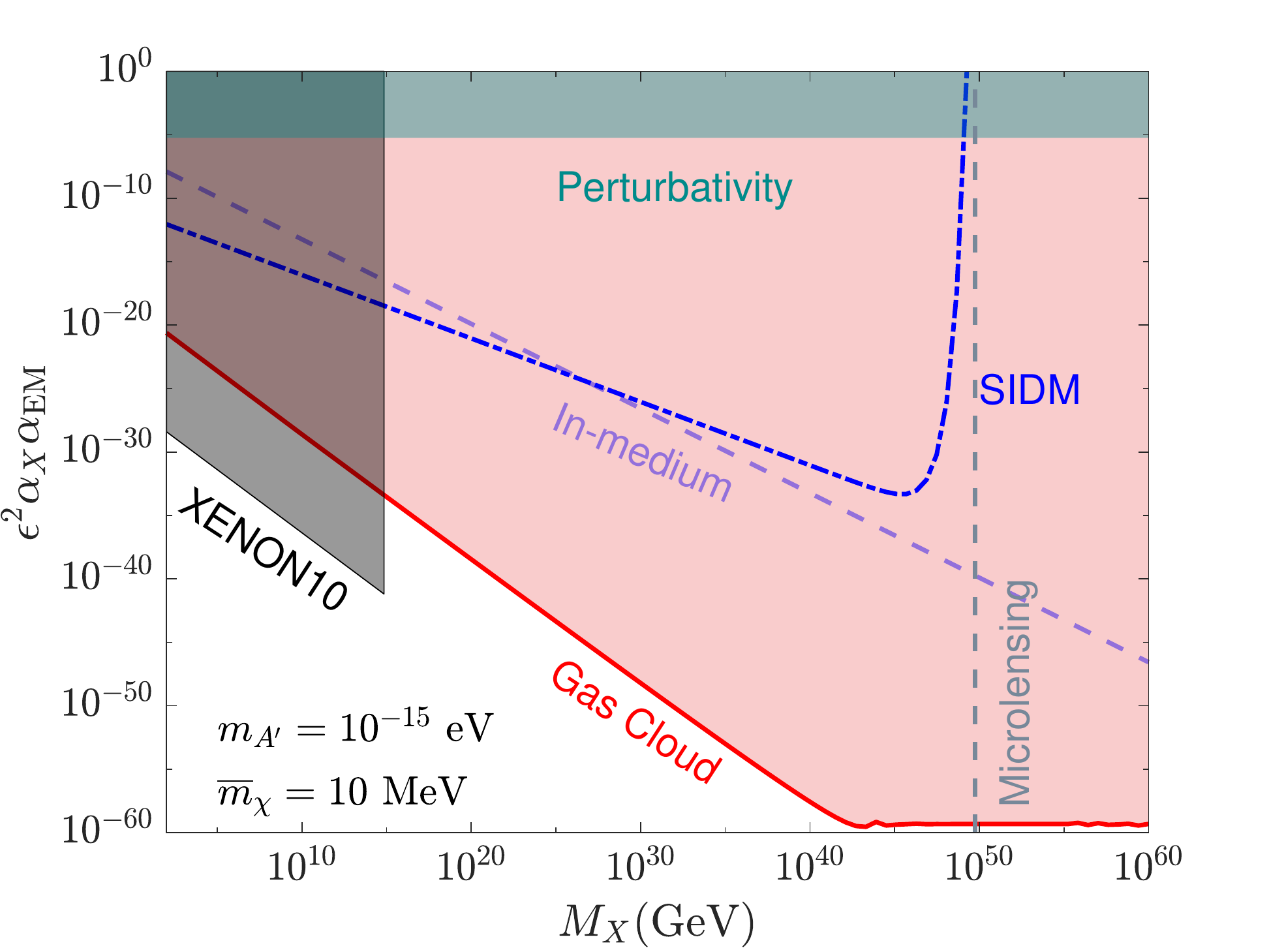}
    \includegraphics[width=0.49\textwidth]{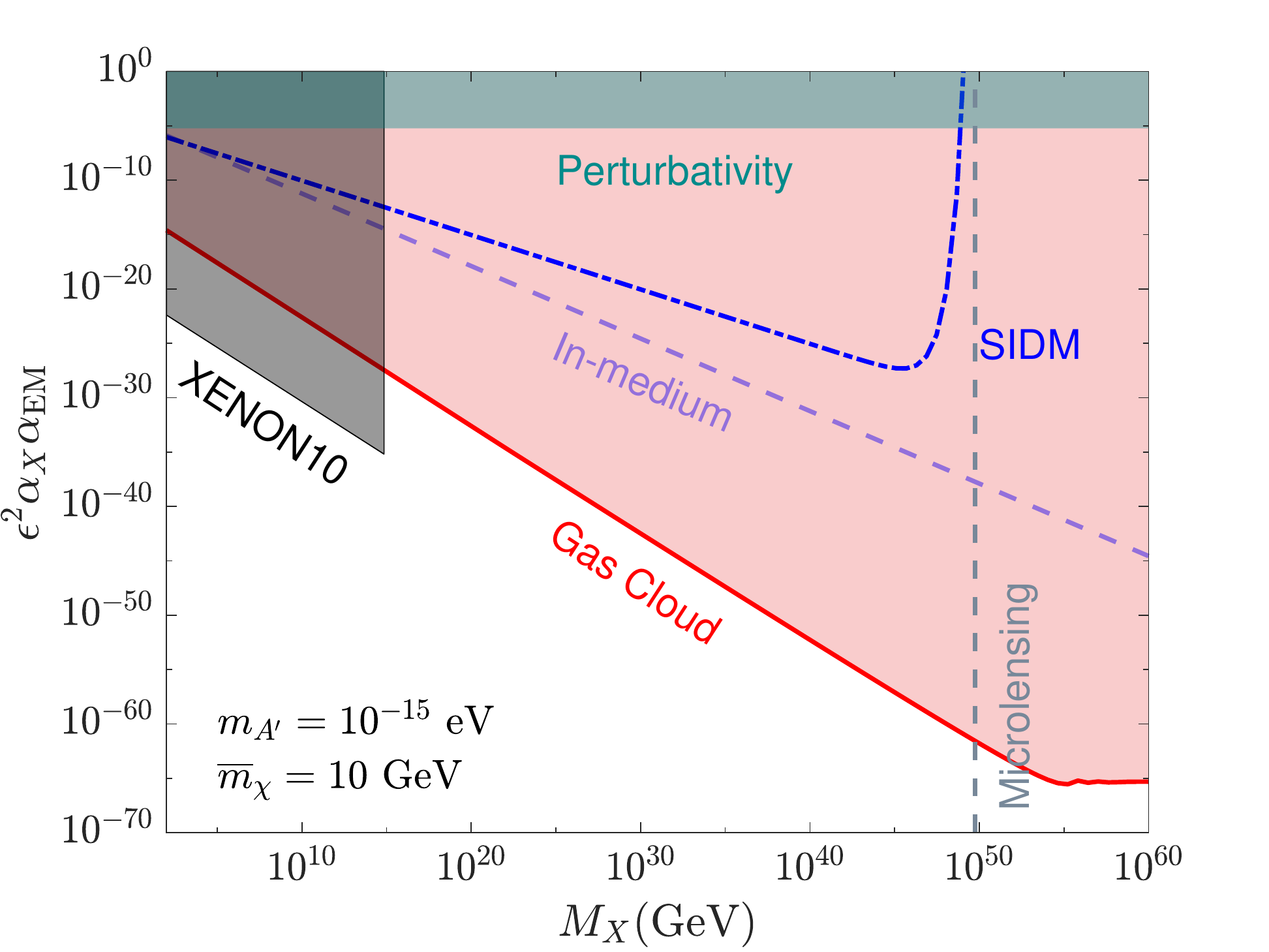}    
    \caption{Constraints are shown for the coupling $\epsilon^2\alpha_X\alpha_{EM}$ of an asymmetric dark matter nugget of mass $M_X$ interacting with SM electrons through a dark photon mediator. The pink region is excluded by cold gas clouds near the Milky Way galactic center. Also shown on the left is the black region excluded by XENON10~\cite{Coskuner:2018are}. The cyan region describes a perturbativity bound which restricts $g_\chi < 1$. The dashed purple and dash-dotted blue lines correspond to in-medium and self interacting dark matter (SIDM) bounds respectively (see text for details). Regions where the Compton wavelength of dark photon is smaller than the nugget size are shaded gray on the right. In the upper two panels we assume the dark photon mass $m_{A'}=10^{-3}$~eV. For the in-medium, perturbativity, and SIDM bounds we take $\epsilon=2\times10^{-9}$, which is the solar lifetime constraint~\cite{Redondo:2015iea}. The lower two panels depict the bounds for $m_{A'}=10^{-15}$~eV. For these panels, in deriving non-gas cloud bounds we fix the mixing parameter $\epsilon=0.1$ to be consistent with Jupiter magnetic field bounds~\cite{Davis:1975mn,Ahlers:2008qc}. Depending on the radius of the ADM composite, microlensing will impose additional constraints when the dark matter mass $M_X\gtrsim 10^{26}$~g~\cite{Croon:2020wpr}, see text for details. This mass threshold is marked by the grey dashed lines.  In the left and right panels we use constituent masses $\mchib=10$~MeV, 10~GeV respectively.}
    \label{fig:darkphotonbound}
\end{figure}

Since we will be most interested here in a light dark photon mediator where $m_{A'}<q$, the dark matter heating rate $dVDHR/dq$ is proportional to $1/q$. The dark matter heating of the gas cloud in this case is dominated by the contribution from low momentum transfer interactions. For the dark matter velocities we are considering, it will be the case that $q<\alpha_\mathrm{EM}m_e$, which is to say electronic ionization of gas cloud atoms is prohibited and only free electrons and ions will contribute to the DM-gas cloud heating rate. Furthermore, it is evident from comparing Eq.~\eqref{eq:heatingratenucleon} and Eq.~\eqref{eq:heatingrateelectron} that heating rate in nuclear scattering is suppressed by a factor of $m_e/m_N$. We have verified throughout the parameter space we consider that the dark matter-gas cloud heating rate is dominated by the nugget scattering with free electrons in the gas cloud for a light dark photon mediator. 

We now turn to the response of the gas cloud medium to long-range interactions with the ADM nugget. The ionized part of the gas clouds can be modeled as a non-relativistic plasma. The effective Compton wavelength of a dark photon in the plasma is described by the Debye length $\lambda_d$ of the plasma, beyond which any electromagnetic interaction is screened. This sets the lower limit of the momentum transfer $q_{\min}=1/\lambda_d$, which also regulates the divergence of the integral in Eq.~\eqref{eq:heatingrateelectron}. In the gas cloud,
\begin{equation}
    q_{\min}=\sqrt{\dfrac{4\pi\alpha_\mathrm{EM}n_e}{T_g}}\,,
\end{equation}
where $T_g$ is the temperature of the gas cloud. The maximum momentum transfer $q_{\max}=2\mu_{Xe}v$ depends on the dark matter velocity explicitly. However, as we mentioned before the dark matter heating rate is dominated by lower momentum transfer. We consequently set $q_{\max}=2m_e v_0$ as the nugget mass $M_X\gg m_e$. We have verified through numerical evaluation of the energy deposition integral that this simplification does not change our results.

Figure~\ref{fig:darkphotonbound} shows gas clouds sensitivity to asymmetric dark matter composites interacting with electrons through a dark photon. For the formalism laid out above, we require the dark photon to mediate a long-range interaction, where the effective Compton wavelength of the dark photon is larger than the size of the dark matter nugget. This implies
\begin{equation}
    \min\left\{\dfrac{1}{m_{A'}}, \lambda_d\right\}>R_X(M_X,\mchib)\,.
\end{equation}
This restriction on the dark photon as a long-range mediator is shown by the shaded gray region in Figure~\ref{fig:darkphotonbound}. Similar to this Debye screening, we also expect the dark photon could have an in-medium correction from the plasma of $\chi$ particles inside the dark matter nugget. Here as in \cite{Coskuner:2018are} we refer to this as the in-medium correction to the dark photon mass $\delta m_{A'}^2\sim \frac{4\pi \alpha_X}{\mchib}\sim 4\pi\alpha_X \mchib^2$. We require that the nugget-SM interaction is not screened by the plasma inside the nugget, i.e.
\begin{equation}
    \dfrac{1}{\sqrt{4\pi\alpha_X}\mchib}>R_X(M_X,\mchib)\,.
\end{equation}
The region where in-medium corrections would need to be accounted for lies above the dashed purple line in Figure \ref{fig:darkphotonbound}. 

A nugget with long-range interactions will also have substantial long-range \emph{self}-interactions. These self-interactions are bounded by observations of the Bullet Cluster~\cite{Feng:2009mn,McDermott:2010pa,Coskuner:2018are}, which restricts $\alpha_X$
\begin{equation}
    \alpha_X \lesssim \frac{1}{4 \pi} \left(\dfrac{\pi}{4}\dfrac{\mchib^4v_{rel}^4}{M_X}\times\dfrac{\mathrm{cm}^2}{g}\right)^{1/2}\exp\left(\frac{1}{2}\sqrt{\dfrac{m_{A'}^2M_X}{\pi}\times\dfrac{\mathrm{cm}^2}{g}}\right)\,.
\end{equation}
This self-interaction constraint is labeled ``SIDM'' with dash-dotted blue lines in Figure~\ref{fig:darkphotonbound}. We also show the region $g_\chi>1$ in cyan where the theory becomes non-perturbative.

When the nugget mass is comparable to the mass of an asteroid or the mass of the sun, the light of background stars can be deflected by dark matter due to gravitational lensing. As is evident from Eq.~\eqref{eq:RX} the radius of a nugget is much smaller than the Einstein radius, consequently microlensing constraints apply for nuggets as point-like lenses when $M_X>10^{26}$~g~\cite{Croon:2020wpr}. This mass scale is marked as dashed grey lines in the lower panels of Figure~\ref{fig:darkphotonbound}, above which ADM nuggets that have radii smaller than the Einstein radius cannot constitute 100$\%$ of dark matter. Given the nugget fraction in dark matter $f_X$, the bound on $\epsilon^2\alpha_\mathrm{EM}\alpha_X$ scales as $1/f_X$. However, care should be taken when applying this bound to ADM nuggets, since in the case of nuggets with an internal vector potential, it is possible to consider ADM nuggets with radii much larger than Eq.~\eqref{eq:RX}, which could have radii much larger than their Einstein radius. We leave exploration of this class of models to future work.

The parameter space explored by gas clouds is enclosed by the pink region in Figure \ref{fig:darkphotonbound}. We see that for a dark photon mass $m_{A'}=10^{-3}$~eV and a constituent mass $\mchib=10$~MeV, gas clouds set a leading bound on dark photon mediated nugget interactions for nugget masses between $10^{17}$~GeV and $10^{27}$~GeV, at which mass the mediator Compton wavelength limit begins restricting interactions. For a larger $\mchib$ the bound would be safe from these Compton considerations up to larger $M_X$, since the dark matter nugget will be more compact for larger constituent masses. The gas cloud bound scales as $1/M_X$ at relatively small nugget mass because of the $N_X^2$ factor arising from the coherent scattering between the nugget and electrons in Eq.~\eqref{eq:dsigmadq}. We see that for some parameters, the bound saturates at a fixed coupling at large enough $M_X$; this is a consequence of suppression from the nugget form factor $|F_X|^2$ when $qR_X>1$ -- for a large enough ADM nugget, the coherent enhancement from constituents collectively coupling to electrons is weakened. On the other hand, the couplings constrained in this case are rather tiny, extending well below $\epsilon^2 \alpha_X \alpha_{EM} = 10^{-60}$.

\section{Composite Dark Matter With Contact Interactions}
\label{sec:contact}

Next we consider a strongly interacting composite dark matter model, where the composite has a fixed cross-section for elastic scattering with nuclei $\sigma_{XN}$, which does not depend on the size or mass of the nucleus. This prescription applies to dark matter that is strongly interacting with nuclei at short range, and is physically larger than any Standard Model nucleus, i.e.~$
    \sigma_{XN} \gg \pi r_n^2$, for nuclear radius $r_n$.
In this case, the differential cross-section for scattering with a nucleus $i$ is given by
\begin{equation}
    \dfrac{d\sigma_i}{dE_R}=\dfrac{m_{N_i}\sigma_{XN}}{2\mu_{XN_i}^2v^2}\,,
    \label{eq:diffcont}
\end{equation}
where $m_{N_i}$ is the mass of the nucleus and $\mu_{XN_i}$ is the dark matter-nucleus reduced mass. Some examples of this sort of dark matter include quark nugget matter \cite{Witten:1984rs} and other large dark matter composite states with strong, short range interactions. In these cases, the dark matter elastically scatters with all nuclei it contacts, and so the total cross-section on any nucleus is equal to the physical area subtended by the dark matter composite.

\begin{figure}
    \centering
    \includegraphics[width=0.8\textwidth]{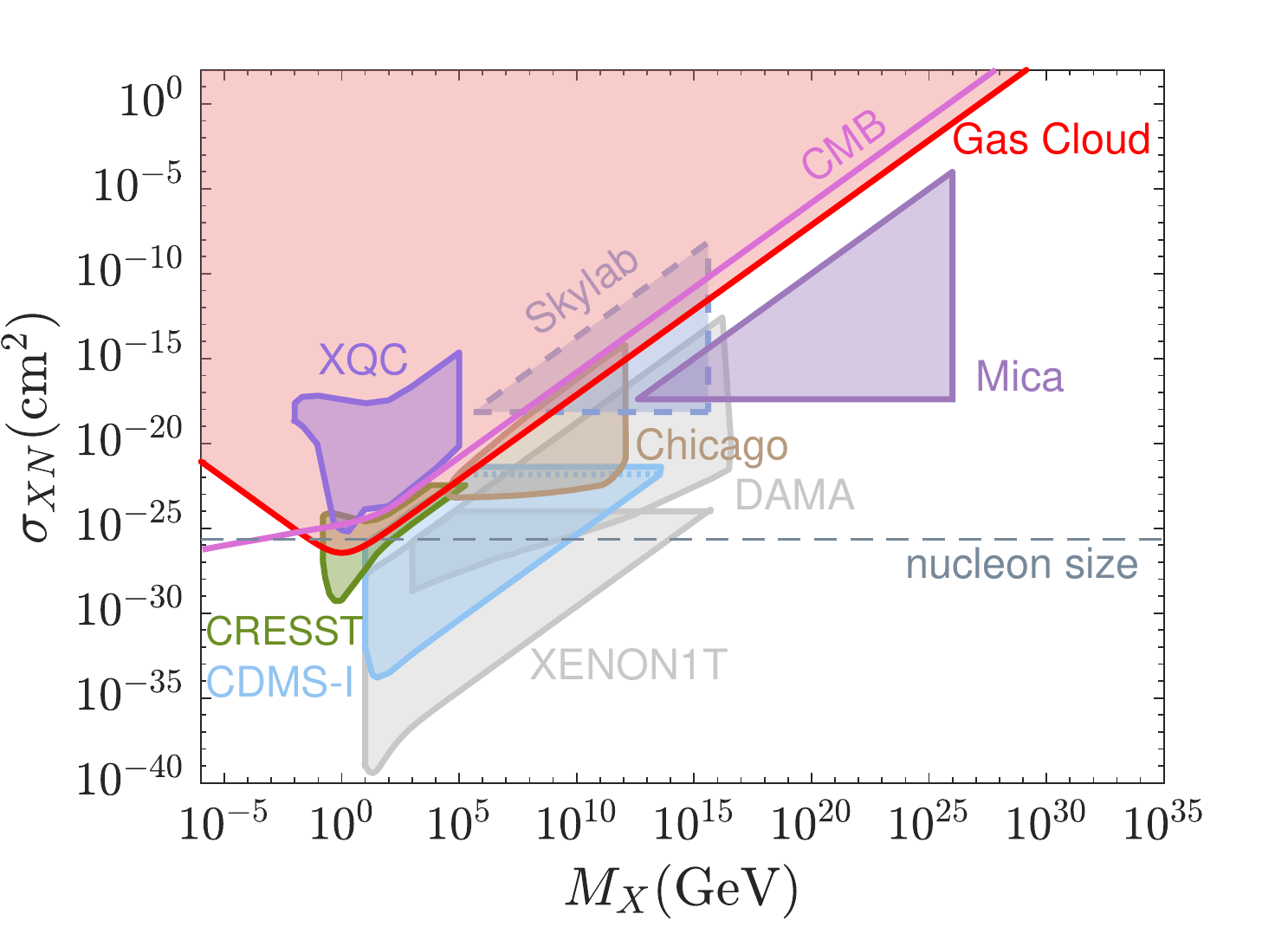}
    \caption{Constraints on large composite dark matter-nuclear contact interaction cross-section. The pink region above the red line is excluded by gas cloud heating. Bounds from CRESST 2017 surface run~\cite{Angloher:2017sxg} and CDMS-I~\cite{Abusaidi:2000wg,Abrams:2002nb} are revisited with a modified version of the code verne~\cite{verne,Kavanagh:2017cru}. The dotted line represents a multiscatter cross-section above which dark matter may scatter multiple times in CDMS-I detectors. The Skylab and Mica bounds are described in Appendix~\ref{sec:appbounds}. Bound from X-ray Quantum Calorimetry Experiment (XQC) is rescaled from above with nitrogen overburden and from below with silicon scattering following Eq.~\eqref{eq:scalingrelation}. The DAMA upper bound is obtained from the overburden of the Earth crust and the lower bound in Ref.~\cite{Bernabei:1999ui} is recast assuming iodine scattering. The top of the XENON1T bound~\cite{Aprile:2018dbl,Clark:2020mna} is determined both by overburden and multiscatter considerations as detailed in the text. A new bound from a shallow-depth experiment carried out at the University of Chicago (here labelled ``Chicago'') is taken from Ref.~\cite{Cappiello:2020lbk}. Constraints from spectral distortions of the cosmic microwave background due to dark matter-nucleon couplings (CMB)~\cite{Dvorkin:2013cea,Gluscevic:2017ywp} are shown in purple. A weaker bound from interstellar gas cooling~\cite{Chivukula:1989cc} is not shown in this figure. Comparable bound can be placed based on the Milky Way satellite observations subject to cosmological and astrophysical assumptions~\cite{Nadler:2020prv}. Other possible constraints could be derived from dark matter induced nuclear transitions~\cite{Lehnert:2019tuw,Song:2021yar}. We have indicated a typical nuclear cross-section with a dashed line labeled ``nucleon size.'' Below this line strongly interacting composite dark matter model must be reconsidered: for such small cross-sections, dark matter cannot elastically scatter with every nucleus it encounters and also be larger than a nucleus.}
    \label{fig:geoxs}
\end{figure}

This can be contrasted with the more common spin-independent nucleon contact interaction formalism, where the nuclear cross-section is associated to a dark matter-nucleon interaction cross-section $\sigma_{Xn}$ through the relation
\begin{equation}
   \sigma_{XN_i}^{(SI)} = \sigma_{Xn}A^2\dfrac{\mu_{X N_i}^2}{\mu_{X n}^2}F_{N_i}^2(q)\,
   \label{eq:scalingrelation}
\end{equation}
We stress that Eq.~\eqref{eq:scalingrelation} will not be used in this analysis, except to translate between prior results obtained with this formula, and the model given by Eq.~\eqref{eq:diffcont}. In the per-nucleon expression, $F_N$ is a nuclear form factor, $A$ is the number of nucleons in the nucleus, and at small momentum transfer $q \ll 1/R_N$, $F_N\simeq 1$. For dark matter much heavier than the nucleus $\mu_{X N}\simeq m_N$, this low momentum transfer limit allows for a simple scaling relation $\sigma_{X N_i}^{(SI)}\simeq \sigma_{X n}A^4$. Eq.~\eqref{eq:scalingrelation} has been employed in most direct detection experiments to obtain the per-nucleon scattering cross-section. However, as suggested by large composite dark matter models \cite{Starkman:1990nj,Jacobs:2014yca,Bramante:2018tos} and notably enunciated in Reference \cite{Digman:2019wdm}, this relation can break down at large cross-section for point-like dark matter particles due to the break down of the first Born approximation and the fact that the cross-section for contact interactions can not be larger the geometric size of the nucleus. On the other hand, exceptions to this argument include composite dark matter and long range interactions between dark matter and nuclei. In both these cases, the interaction between dark matter and the nucleus can have a super-nuclear sized cross-section. In Section \ref{sec:ac} we have considered the case where dark matter composite interactions are mediated by a long-range force. Here we investigate the simple possibility stated above (Eq.~\eqref{eq:diffcont}), that a super-nuclear sized composite dark matter particle interacts elastically with all nuclei it contacts. In this case dark matter is opaque to nuclei, so that the scattering cross-section is only determined by the geometric size of the composite state, and the same cross-section applies to all nuclei.

Gas clouds will have special sensitivity to large composite dark matter that scatters elastically with nuclei. As detailed in References \cite{Bhoonah:2018wmw,Bhoonah:2018gjb}, gas clouds have the unique advantage of being extremely large detectors, that are sensitive to thermal energy dark matter can deposit. In gas clouds, the volumetric heating rate from the dark matter contact interaction given above is
\begin{equation}
    VDHR=\sum\limits_i n_X n_{i}\sigma_{X N}\langle v E_R\rangle\,,
\end{equation}
where the sum runs over all nuclei in gas clouds and $n_i$ is the number density of the nuclei. The average recoil energy $\langle E_R\rangle=\mu_{X N}^2v^2/m_N$, which implies
\begin{equation}
        VDHR=\sum\limits_i\dfrac{\rho_X}{M_X}f_i\dfrac{m_n n_b}{m_{N_i}^2}\mu_{X N_i}^2\sigma_{X N}\int v^3f(v)\,.
    \label{eq:heatingratecontact}
\end{equation}

In Figure~\ref{fig:geoxs}, by requiring the dark matter heating rate not exceed the gas cloud cooling rate for some particularly cold gas clouds near the galactic center, we place constraints on dark matter's geometric cross-section with nuclei. Here we have used all the same parameters, provisos, and procedures laid out in Section \ref{sec:ac}. In addition, we have performed a good amount of additional analysis, in order to include additional bounds on this parameter space from direct detection experiments on earth, along with Skylab and Mica bounds detailed in Appendix \ref{sec:appbounds}. 

In the remainder of this Section, we discuss our procedure for obtaining bounds from a number of direct detection experiments. We have rescaled the X-ray Quantum Calorimeter (XQC) rocket bound on dark matter for the contact interaction model of this paper, using Eq.~\eqref{eq:scalingrelation}. The XQC experiment has an overburden comes from atmosphere. So we set $A\simeq 14$ when recasting the upper bound obtained in \cite{Erickcek:2007jv} while assuming $F_N=1$. Dark matter interacts primarily with the Silicon targets in the XQC detector so we assume $A\simeq 28$ to recast the XQC lower bound on per nucleon scattering. 

We obtain CRESST \cite{Angloher:2017sxg} and CDMS-I \cite{Abusaidi:2000wg,Abrams:2002nb} bounds on strongly interacting composite contact interaction dark matter with a modified version of the verne code~\cite{Kavanagh:2017cru}. We have used the verne code to compute dark matter particle energy loss on its path through Earth and structural overburdens, to obtain an accurate spectrum of dark matter particle energies at CRESST and CDMS-I.  The average loss of kinetic energy when dark matter traverses the overburden can be approximated as 
\begin{equation}
    \dfrac{d\langle E_\chi\rangle}{dt}=-\sum\limits_i n_i\langle E_R\rangle_i \sigma_i v\,,
    \label{eq:energyloss}
\end{equation}
where $E_R$ is the nuclear recoil energy and the sum runs over all nuclei in the overburden. 
After traveling a distance $D$, the change in the dark matter velocity is obtained from Eq.~\eqref{eq:energyloss}
\begin{equation}
    \dfrac{dv}{dD}=-\dfrac{\sigma_{XN}v}{M_X}\sum\limits_i \dfrac{\mu_{XN_i}^2n_i}{m_{Ni}}\,.
    \label{eq:dvdD}
\end{equation}
The overburden modifies the final velocity distribution of dark matter reaching the detector, and the nuclear scattering rate in the detector is expressed as
\begin{equation}
    R=n_T\dfrac{\rho_\chi}{M_\chi}\int_{v_{\min}} dv vf_\mathrm{det}(v) \int_{E_{R,{\min}}}^{E_{R,{\max}}} \dfrac{d\sigma}{dE_R} dE_R\,,
    \label{eq:eventrate}
\end{equation}
where the dark matter velocity distribution upon arrival $f_\mathrm{det}(v)$ differs from the initial Maxwellian distribution. For each experiment, we use the threshold energy to define $v_{\min}$, the minimum dark matter velocity necessary to trigger the detector after traversing the overburden. Using this, an upper limit to the excluded region is determined by requiring the number of nuclear recoil events in the detector be less than the events observed. This procedure was undertaken for both the CRESST and CDMS-I detectors, for which we used recoil energies of $20-600$ eV and $10-100$ keV, respectively.

If the contact cross-section is large enough, dark matter will scatter multiple times in a dark matter detector \cite{Bramante:2018qbc,Bramante:2018tos}. Our multiscatter treatments for CRESST and CDMS-I are as follows. The CRESST detector in the 2017 Sapphire crustal surface run configuration consisted of a cooled $(5~\mathrm{mm})^3$ Al$_3$O$_2$ crystal \cite{Angloher:2017sxg}. Dark matter with a cross-section larger than $\sigma_\mathrm{th}=3.5\times10^{-23}~\mathrm{cm}^2$ will be able to scatter more than twice and deposit more than 600~eV nuclear recoil energy in the crystal. Consequently we truncate the CRESST exclusion region at this multiscatter threshold cross section.

To estimate multiscatter dark matter bounds for CDMS-I, we first characterize the physical dimensions of the detector. A single CDMS-I detector (BLIP) consists of a cylindrical crystal of high-purity germanium with 6~cm diameter and 0.3~cm thickness. Three such detectors were employed in the CDMS-I shallow site search \cite{Abusaidi:2000wg,Abrams:2002nb}. A threshold cross-section $\sigma_\mathrm{th}=1.5\times 10^{-22}~\mathrm{cm}^2$ is required for dark matter to scatter twice in these detectors. This multiscatter threshold is marked with the blue dotted line in Figure~\ref{fig:geoxs}. We will find that we can obtain a bound that extends slightly into this multiscatter region, by requiring that prospective dark matter events saturate the number of candidate events reported in \cite{Abrams:2002nb}. However, our approach will be limited by the fact that \cite{Abrams:2002nb} did not report detailed data for events with nuclear recoil energies exceeding 100 keV. 

In more detail, the BLIP detectors were covered with a 4.1~cm-thick plastic scintillator as muon veto, with a veto threshold of 2.6 MeV. There were 27 single scatter candidate events that passed the muon veto, and 3287 muon-coincident (vetoed) ``neutron candidate multiscatter events'' \cite{Abrams:2002nb} that failed the muon veto. Therefore it would be conservative to assume multiscatter dark matter failed the muon veto if it deposits enough energy in the scintillator to trigger the veto. (To conservatively limit our bound, we could assume multiple scattering dark matter nuclear scattering interactions can efficiently deposit all their recoil energy in scintillator -- in fact nuclear scattering in scintillators is usually less efficient at creating detectable photoelectrons than scattering of charged particles like muons \cite{Bramante:2018tos}.) However, our analysis will not have to consider the muon veto, because using a typical scintillator composed of $^{12}$C with 1~g/cm$^3$ density, dark matter would need a cross-section larger than $\sim 4.2\times 10^{-21}~\mathrm{cm}^2$ to trigger this muon veto. Nevertheless, well below this cross-section, dark matter is already depositing more than 100 keV in a single passage through a BLIP detector, at a cross-section $\sim 4.2\times10^{-22}~\mathrm{cm}^2$. As already mentioned, beyond recoil energies of 100 keV, multiscattering data is not available in the CDMS-I publication \cite{Abrams:2002nb}. So we truncate our CDMS-I multiscattering bound at the 100 keV detector energy deposition cross-section $4.2\times10^{-22}~\mathrm{cm}^2$, which lies below the muon veto threshold cross-section. In other words, because we do not have high energy CDMS-I event data, our analysis can only consider events that did not trigger the muon veto. Throughout our CDMS-I sensitivity region, we require that $27$ dark matter particles cross the detector in 99.4 live days. This limits the dark matter mass range to below $4\times 10^{13}$~GeV. A future analysis which has access to the number of muon-coincident multiscattering events with BLIP detected energies exceeding 100 keV, could bound larger nuclear scattering cross-sections, utilizing high energy events that did not pass the muon veto.

To derive the XENONT1T upper bound we can rewrite Eq.~\eqref{eq:dvdD} as
\begin{equation}
    \dfrac{dE}{dD} = -2 \dfrac{\sigma_{XN}E}{M_X}\sum\limits_i \dfrac{\mu_{XN_i}^2n_i}{m_{Ni}}\,.
\end{equation}
The cross-section upper bound can be estimated by requiring the dark matter particles with maximum kinetic energy $E_{0,\max}$ barely triggers the detector threshold after traversing the overburden of length $L$, i.e.
\begin{equation}
    \sigma_{XN}=\dfrac{M_X}{2L}\ln({E_{0,\max}}/{E_{f,\min}})\left(\sum\limits_i \dfrac{\mu_{XN_i}^2n_i}{m_{Ni}}\right)^{-1}\,,
    \label{eq:xenon1tupper}
\end{equation}
where $E_{f,\min}$ is the minimum dark matter kinetic energy upon reaching the detector. XENON1T experiment~\cite{Aprile:2018dbl} was carried out at  the INFN Laboratori Nazionali del Gran Sasso (LNGS) about 1400 meters underground. We assume the Earth crust consists of (O, Si, Al, Fe, Ca, Na, K, Mg) with the mass fraction (0.466, 0.277, 0.081, 0.050, 0.036, 0.028, 0.026, 0.015). The detector threshold for nuclear recoil is 4.9~keV. The threshold cross-section for dark matter to scatter twice in the meter-scale detector is $\sigma_\mathrm{th}=1.4\times 10^{-24}~\mathrm{cm}^2$, above which XENON1T loses its constraining power. The threshold cross-section $\sigma_\mathrm{th}$ is combined with the overburden to set the upper limit in the cross-section exclusion region. As with XQC, the XENON1T lower bound is rescaled from per-nucleon cross-section to contact interaction cross-section with xenon nuclei. We note that our results appear in good agreement with a recent per-nucleon bound obtained in \cite{Clark:2020mna}.

In 1998, the DAMA collaboration constructed two co-planar configurations of sodium iodide crystals to search for strongly interacting dark matter, including dark matter that would interact multiple times in these detectors~\cite{Bernabei:1999ui}. Similar to XENON1T, we use Eq.~\eqref{eq:xenon1tupper} to set the overburden upper bound by taking into consideration the major elements in the Earth crust and by assuming 4~keV detector threshold. We have checked that this prescription for the overburden upper bound, which in the case of the dark matter-nucleon scattering cross section is $\sigma_{Xn}=\frac{M_X}{2L}\ln({E_{0,\max}}/{E_{f,\min}})\left(\sum\limits_i \frac{\mu_{XN_i}^4n_iA_i^2}{\mu_{Xn}^2 m_{Ni}}\right)^{-1}$, yields a slightly more conservative bound than reported in~\cite{Bernabei:1999ui}. In rescaling the \cite{Bernabei:1999ui} lower bound, we assume dark matter scatters predominantly with iodine, which is the heaviest nucleus in the detector and will yield the most conservative bound.

\section{Conclusions}
\label{sec:conc}
We have found that interstellar gas clouds can be re-purposed as exquisitely sensitive calorimetric detectors in the hunt for heavy composite dark matter. In particular, cold gas clouds are sensitive to models of composite dark matter that interact through a long-range force, or through a contact interaction with a cross-section in excess of a nuclear cross-section. Here we have focused on constraints from cold gas clouds identified near the galactic center, which span tens of parsecs and have lifetimes spanning over a hudnred million years. These gas clouds are thermally sensitive to dark matter interactions for masses extending to $10^{60}$ GeV. In the case of long-range interactions, we focused on a particular model of asymmetric composite dark matter coupled to Standard Model particles through a dark photon mediator. We also considered heavy composite dark matter with a super-nuclear size contact interaction cross-section with nuclei. In both cases we found that cold gas clouds about a kiloparsec from the galactic center appear to place leading bounds on composite dark matter, assuming a standard Milky Way dark matter halo density profile.

In addition to our work investigating gas clouds as composite dark matter detectors, we have also recast bounds from experiments like CDMS, CRESST, DAMA, and XQC on strongly interacting composite dark matter that has a fixed contact interaction with nuclei. In the case of our CDMS-I re-analysis, the treatment was necessarily rudimentary and conservative -- in order to make sure our bounds were not overstated, we treated every event in CDMS-I BLIP detectors as a potential multiscatter event. Stronger bounds could be obtained in the future with a more complete analysis of the full CDMS-I dataset.

Much about the thermal properties of interstellar gas clouds remains to be explored in future work. In our treatment of gas cloud sensitivity to dark matter here, we attributed all gas cloud heating to dark matter heating in order to set conservative bounds. In doing so, we neglected gas cloud heating contributions from cosmic rays, starlight, and other galactic sources that would tend to inject more energy into gas clouds. Incorporating realistic values for these heating sources would further limit the heating contribution from dark matter, resulting in a tighter bound on dark matter couplings. However, this requires careful modeling of gas cloud characteristics, whilst marginalizing over gas cloud parameters. We address these gas cloud calibration considerations in forthcoming work.

\section*{Acknowledgements}
The work of JA, JB, SS, NS is supported by the Natural
Sciences and Engineering Research Council of Canada (NSERC). We thank Levente Balogh, Fatemeh Elahi, Matt Leybourne, Nirmal Raj, and Aaron Vincent for useful discussions. Research at Perimeter Institute is supported in part by the Government of Canada through the Department of Innovation, Science and Economic Development Canada and by the Province of Ontario through the Ministry of Colleges and Universities.

\appendix
\section{Bounds from Experiments Near Earth}
\label{sec:appbounds}

\indent {\em Mica:} An experiment looking for tracks in ancient mica set a bound on a cosmogenic particle flux $\Phi_{mica} < 10^{-18} ~{\rm cm^{-2} ~s^{-1} ~sr^{-1}} $ \cite{Price:1986ky}. Assuming a local dark matter density of $\rho_X = 0.4 ~{\rm GeV/cm^3}$, this implies a bound on dark matter particles for masses $M_X \lesssim 10^{26} ~{\rm GeV}$. The energy deposition threshold at which etchable tracks were detectable in mica is given by $\rho_{mica}^{-1} dE/dx > 2.4 ~{\rm GeV ~cm^2~g^{-1}}$. Reminding ourselves that the energy deposition rate for a strongly interacting composite will be $\rho_{mica}^{-1} dE/dx = \sigma_{XN} v^2$, this corresponds to a threshold scattering cross-section $\sigma_{XN} \simeq 4 \times 10^{-18}~{\rm cm^2}$. To determine the overburden, we conservatively set the cross-section at which the Earth's crust would slow the DM and prevent it from leaving an etchable track in mica, as the cross-section at which 50\% of the DM's kinetic energy would be lost by transiting three kilometers of crust. This sets an overburden cross-section $\sigma_{XN} = {\rm Log}[E_i/E_f] M_X/(\rho_{cr} \ell_{cr})$, where we take $\rho_{cr} \approx 3 ~{\rm g/cm^3}$ and $\ell_{cr} = 3~{\rm km}$ as the density and span of the Earth's crust above the mica, and we take $E_i/E_f < 2$. This restricts mica bounds to cross-sections $\sigma_{XN} <  10^{-15}~{\rm cm^2}~({M_X/10^{15}~\rm GeV})$.
\\

\noindent {\em Skylab:} In upcoming work, we address plastic etch bounds on dark matter in greater detail \cite{Bhoonah:2020fys}; here we provide a conservative bound and brief summary. One of many experiments conducted on-board the Skylab space station was the study of the high $Z$ composition of cosmic rays \cite{Shirk1978,Starkman:1990nj}. The detector used was a 1.17 m$^{2}$ array of $36$ modules of Lexan plastic track detectors, each containing thirty-two $250 \ \mu$m thick sheets. Over a data collection period lasting 253 days, about 150 events were observed to have passed through all sheets. Assuming conservatively that all of these were dark matter particles interacting with the detector, we can constrain dark matter for masses $M_{X} \lesssim 4 \times 10^{15} $\ GeV. The Lexan (L) detector threshold given in  \cite{Shirk1978} is $\rho^{-1}_{L} \frac{dE}{dx} = 0.4 $\ GeV cm$^{-1}$g$^{-1}$. For this threshold, the threshold scattering cross-section is $\sigma_{XN} \simeq 7\times 10^{-19}$ cm$^{2}$. To set the overburden bounds, we require that the dark matter passing through the $1$ g\,cm$^{-2}$ aluminium wall against which the plastic track detectors are mounted retain $90\%$ of their energy, giving an overburden constraint $\frac{\sigma_{XN}}{M_{X}} \lesssim 1 $\ g$^{-1}$ cm$^{2}$. Our overburden constraints here are similar to those obtained in \cite{Starkman:1990nj}, which imposed a $90\%$ energy retention criterion, but for dark matter passing through 0.25 cm of Lexan, which translates to $\frac{\sigma_{XN}}{M_{X}} \lesssim 3 $\ g$^{-1}$cm$^{2}$. 

\bibliographystyle{JHEP.bst}

\bibliography{composite.bib}

\end{document}